\documentclass[reqno]{amsart}
\usepackage{a4wide,array}

\numberwithin{equation}{section}

\newtheorem{lemma}{Lemma}

\theoremstyle{definition}
\newtheorem*{ack}{Acknowledgements}

\newcommand\dv{D}
\newcommand\Tr{\mathrm{Tr}\,}

\begin{document}
\title{A Spinorial Hamiltonian Approach to Gravity}
\author{James D.E. Grant}
\address{Department of Mathematics\\
University of Hull\\ Hull HU6 7RX\\ U.K.}
\begin{abstract}
We give a spinorial set of Hamiltonian variables for
General Relativity in any dimension greater than $2$. This approach
involves a study of the algebraic properties of spinors in higher
dimension, and of the elimination of second-class constraints from
the Hamiltonian theory. In four dimensions, when restricted to the
positive spin-bundle, these variables reduce to the standard
Ashtekar variables. In higher dimensions, the theory can either be
reduced to a spinorial version of the ADM formalism, or can be left
in a more general form which seems useful for the investigation of
some spinorial problems such as Riemannian manifolds with reduced
holonomy group. In dimensions $0 \pmod 4$, the theory may be recast
solely in terms of structures on the positive spin-bundle
$\mathbb{V}^+$, but such a reduction does not seem possible in
dimensions $2 \pmod 4$, due to algebraic properties of spinors in
these dimensions.
\end{abstract}
\maketitle

\section{Introduction}

One of the central elements of the Ashtekar variables approach to
canonical gravity \cite{A1} is the projection, to a $3$-dimensional
hypersurface, of the natural connection on the positive spin-bundle
$\mathbb{V}^+$ of a four manifold. This connection, originally
introduced by Sen \cite{S2}, contains information about both the
$3$-dimensional spin-connection, and the extrinsic curvature of the
hypersurface in a way which leads to a considerable simplification
of the constraints of the Hamiltonian version of the theory. If we
consider the Riemannian version of the theory, then Ashtekar's
approach is very much based on the fact that the four-dimensional
spin-group Spin$_4$ is not simple, but decomposes as $\mathrm{SU}
(2) \times \mathrm{SU} (2)$. This decomposition means that the
spin-connection decomposes into two independent $\mathrm{SU}(2)$
connections on the positive and negative chirality spin-bundles
$\mathbb{V}^{\pm}$. (Similarly, there exists a reformulation of
$3$-dimensional gravity in $\mathrm{SU}(2)$ spinor form.) However,
no such decomposition happens in higher even dimensions, with the
connections on $\mathbb{V}^{\pm}$ in higher dimensions carrying all
of the information of the full spin-connection. In odd dimensions,
there is no chiral decomposition of the spin-bundle at all. It
therefore seems that Ashtekar's approach is very much limited to
$3$ and $4$-dimensional spaces.

Independently, however, Witten
introduced a similar spinor connection in
his proof of the Positive Energy
Theorem \cite{W1}. Although his argument is
motivated by supergravity
considerations, it is independent of the
dimension of the spacetime (all
that is required is that the
hypersurfaces we consider admit
a spin structure), and works equally
well whether we work on the full
spin-bundle or, in even dimensional
spacetimes, restrict to the positive
or negative spin-bundle.

The question we wish to analyse
is whether one can develop a Hamiltonian
theory based on Witten's connection
in any dimension. Based on a
(non-chiral) generalisation of the
action introduced in four dimensions
to describe Ashtekar's theory
\cite{JS,S1}, we construct a Hamiltonian
theory in any dimension (greater
than $2$) which reduces to Ashtekar's
theory in dimension $4$ when
restricted to the positive spin-bundle. In
general, setting up the theory
requires an analysis of the algebraic
structures on spin-bundles in
general dimension, and the Hamiltonian
theory contains extra constraints
and variables which do not naturally
appear in the standard
$4$-dimensional theory.
Many of these constraints
and variables can be systematically
removed from the theory, and reduce
the theory to a minimal version
which is independent of dimension. This
theory contains both first and
second-class constraints, but in
dimension $3$, or dimension $4$
restricted to $\mathbb{V}^+$, all of the
second-class constraints drop out
of the theory, and the theory reduces
to the relevant version of the
Ashtekar theory. In all other cases, we
can remove the second-class
constraints by Dirac's procedure, leading to
a theory with only first-class
constraints. It is sometimes advantageous
to work with the resulting formalism
directly, but alternatively one can
remove some of the first-class
constraints and reduce the theory to a
spinorial version of the ADM formalism.

The plan of the paper is as
follows. We begin by explaining the
spinorial action principle we
will use. To begin with, this is simply a
generalisation of the work of
\cite{JS,S1}. However, in higher dimensions,
there are some subtleties with
the equations of motion (in particular in
dimensions $2 \pmod 4$) which
require a careful study of the properties
of spin-bundles and Clifford
algebras in different dimensions. We then
proceed, in Section~\ref{sec:hamdec},
with the Hamiltonian decomposition of the
theory. Although, again, part of
this work is standard, we find that the
theory contains new constraints
and variables in higher dimensions, with
new sets of each appearing in
dimensions $2 \pmod 4$ or $3 \pmod 4$
depending on the approach one
adopts. In Section~\ref{sec:higher},
we show how these extra
constraints may be unravelled and removed
from the theory, along with
the extra variables. For completeness,
it is then shown, in
Section~\ref{sec:3and4}, how the
resulting minimal theory may be reduced
to the known Ashtekar version of
the $3$ and $4$ dimensional theory.
Finally, we consider the general
version of this minimal theory in
Section~\ref{sec:general}.
We point out some circumstances
in which it is useful to work
with this theory directly, notably
in the study of Riemannian metrics with
reduced holonomy group. However,
we also show how, if desired,
the theory can be reduced to a
spinorial analogue of the
orthonormal-frame approach
to the ADM formalism.

It should be noted that in
higher even dimensions the natural
connections on $\mathbb{V}^{\pm}$
are not independent, and carry all of
the information of the space-time
connection. It is therefore not to be
expected that there will be any
particular simplification in looking
at a chiral version of the theory.
In dimensions $0 \pmod 4$, it turns
out that it is still possible to
rewrite the theory simply in terms of
the connection on $\mathbb{V}^+$,
say, and it may be the case that such
an approach would be useful if
we were to couple the theory to chiral
fermions. In dimensions $2 \pmod 4$,
for algebraic reasons it does not
seem possible to reduce the theory
to $\mathbb{V}^+$. These algebraic
arguments are essentially
the same as those which suggest that coupling
chiral fermions to gravity
leads to formidable problems with the
quantisation of the theory
in these dimensions \cite{AW}.

\section{Connections and Curvature}

We work on a real spin manifold $X$ of dimension $n \ge 3$. We
assume that $X$ carries a pseudo-Riemannian metric $\mathbf{g}$ of
signature $(r, s)$ and that, locally, we may introduce a
pseudo-orthonormal basis $\{ \boldsymbol{\epsilon}^A | A = 1 \dots
n \}$ for the cotangent bundle $T^*X$ in terms of which the metric
may be written
\[
\mathbf{g} = {\eta}_{AB}
\boldsymbol{\epsilon}^A \otimes
\boldsymbol{\epsilon}^B,
\]
where the matrix ${\eta}_{AB}$ takes the diagonal form
\[
{\eta}_{AB} = \mathrm{diag}
[\underbrace{1 \dots 1}_{r},
\underbrace{-1 \dots -1}_{s}].
\]
(Generally upper case letters $A, B, \dots$ will denote
internal $\mathrm{SO}(r, s)$ indices whilst lower case
letters $a, b, \dots$ will denote space-time coordinate
indices. Similar conventions will be assumed for
spatial indices, when we later consider Hamiltonian
decompositions.)
The spin connection $\Gamma$ on the (pseudo)-orthonormal
frame bundle is uniquely determined by torsion-free
condition that the frame $\boldsymbol{\epsilon}$ is
covariantly closed
\[
d^{\boldsymbol{\Gamma}} \boldsymbol{\epsilon} = 0.
\]
The fact that $\Gamma$ is a connection on the (pseudo)-orthonormal frame
bundle means that the connection automatically annihilates the metric
\[
\nabla \mathbf{g} = 0.
\]
We are free to make internal SO$(r, s)$ transformations of the form
\[
\boldsymbol{\epsilon} \mapsto
{\Lambda} \boldsymbol{\epsilon},
\]
where
\[
\Lambda = \mathrm{exp}
\left( \frac{1}{2} {\alpha}_{AB} M^{AB} \right),
\]
and the generators $M^{AB}$ form a representation of
the Lie algebra of SO$(r, s)$:
\begin{equation}
\left[ M^{AB}, M^{CD} \right] =
- {\eta}^{AC} M^{BD} + {\eta}^{AD} M^{BC} +
{\eta}^{BC} M^{AD} - {\eta}^{BD} M^{AC}.
\label{sorsalg}
\end{equation}

The complex Clifford algebra $\mathbb{C}\mathrm{l}_n$ is an algebra
over $\mathbb{C}$, with identity $\mathrm{Id}$, generated by $T^*
X$, and may be viewed as the algebra generated by the
skew-symmetrised products of objects $\gamma^A$ which obey the
relation
\begin{equation}
{\gamma}^A {\gamma}^B +
{\gamma}^B {\gamma}^A = - 2 {\eta}^{AB} \mathrm{Id}.
\label{gamalg}
\end{equation}
The Clifford algebra over $T^* X$ is canonically isomorphic as a
vector space to the exterior algebra $\Lambda^* X$, so given any
differential form $\boldsymbol{\lambda}$ on $X$, we may consider
the corresponding section of the Clifford algebra bundle, denoted
$\boldsymbol{\sigma} (\boldsymbol{\lambda})$. In particular, we may
define $\gamma^A =
\boldsymbol{\sigma} ( \boldsymbol{\epsilon}^A )$.
These objects may then be viewed as sections of the bundle
$\mathrm{End} \mathbb{V}$ of endomorphisms of the
spin-bundle $\mathbb{V}$.

The generators of the spin-$1/2$ representation of the
$\mathfrak{so}(r, s)$ algebra~\eqref{sorsalg} are
\[
{\Sigma}^{AB} = - \frac{1}{4}
\left[ {\gamma}^A,{\gamma}^B \right].
\]
The natural spinorial covariant
derivative of a spinor field $\psi$ is
defined in terms of the image in
the Clifford algebra of the spin
connection $\boldsymbol{\Gamma}$ by
\begin{eqnarray*}
{\nabla} \psi &=& d \psi +
\frac{1}{2} \boldsymbol{\Gamma}_{AB} {\Sigma}^{AB} \psi
\\
&=& d \psi + \mathbf{A} \psi, \qquad
\forall \psi \in \Gamma(\mathbb{V}),
\end{eqnarray*}
where
\[
\mathbf{A} := \frac{1}{2} {\Gamma}_{AB} {\Sigma}^{AB}
\]
is the spinor connection. The curvature of this connection,
$\mathbf{F}$, is defined by the relation
\[
\left( \left[ \nabla_{\mathbf{X}}, \nabla_{\mathbf{Y}} \right]
- \nabla_{\left[ \mathbf{X}, \mathbf{Y} \right]} \right) \psi
=
\mathbf{F}_{\mathbf{X} \mathbf{Y}} \psi,
\qquad \forall \psi \in \Gamma ( \mathbb{V} ),\qquad
\forall \mathbf{X}, \mathbf{Y} \in \Gamma (TX).
\]
Defining the curvature of the spin connection
\[
\mathbf{R} = d \boldsymbol{\Gamma} +
\frac{1}{2} \left[ \boldsymbol{\Gamma},
\boldsymbol{\Gamma} \right],
\]
we can then identify the curvature
$\mathbf{F}$ with its spinorial image
\begin{equation}
\mathbf{F}_{\mathbf{X} \mathbf{Y}} =
\frac{1}{2} \mathbf{R}_{\mathbf{X} \mathbf{Y}AB} {\Sigma}^{AB},
\qquad
\forall \mathbf{X}, \mathbf{Y} \in \Gamma ( TX ).
\label{FR}
\end{equation}

Standard $\gamma$-matrix techniques, and the cyclic identity for
the Riemann tensor yield the identity
\[
R_{abAB} \, {\gamma}^b {\gamma}^A {\gamma}^B
= - 2 \, r_{ab} {\gamma}^b,
\]
where $r$ denotes the Ricci tensor of
the metric $g$ and we have defined
the spacetime $\gamma$-matrices
\[
{\gamma}^a := {\gamma}^A {\epsilon}_A{}^a.
\]
This relation in turn implies that
\[
R_{abAB} \, {\gamma}^a {\gamma}^b
{\gamma}^A {\gamma}^B = - 2 \, s \, \mathrm{Id},
\]
where $s = \Tr \ r$ is the scalar curvature
of the metric $\mathbf{g}$. Therefore
\begin{align*}
s &= - \frac{1}{2 {\dv}} \Tr
\left(
R_{abAB} {\gamma}^a {\gamma}^b {\gamma}^A {\gamma}^B
\right)
\\
&= \frac{2}{\dv} \Tr
\left( F_{ab} {\gamma}^{ab} \right),
\end{align*}
where we have define the skew-symmetrised product of
$\gamma$-matrices
\[
{\gamma}^{a_1 \dots a_p} = \frac{1}{p!}
\left[ {\gamma}^{a_1} \dots {\gamma}^{a_p}
\pm {\mbox{even and odd permutations}}
\right],
\]
and $\dv = \mathrm{dim} \mathbb{V}$
is the rank of the spin bundle.

We can therefore rewrite the Einstein Hilbert action as
\begin{align*}
S_{EH} &= \frac{1}{16 \pi G} {\int}_X g^{1/2} \, s \, d^n x
\\
&= \frac{1}{8 \pi G \, {\dv}} {\int}_X g^{1/2}
\, \Tr
\left[ F_{ab} {\gamma}^{ab} \right] \, d^n x.
\end{align*}
It will be useful to define units in which
\[
4 \pi G \, {\dv} = 1
\]
in which case we have
\begin{equation}
S_{EH} = \frac{1}{2} {\int}_X g^{1/2}
\, \Tr
\left[ F_{ab} {\gamma}^{ab} \right]\, d^n x.
\label{ehaction}
\end{equation}

This action has been considered
in the special case of
dimension $4$ with the connection
restricted to the positive chirality
spin bundle in connection with Ashtekar
variables \cite{JS,S1}.

\subsection{Equations of Motion}
\label{sec:eom}

Consider now the equations of motion that follow from the
action~\eqref{ehaction}. We take the connection $\mathbf{A}$ and
the spacetime $\gamma$-matrices $\gamma^a$ as the independent
variables with the inverse spacetime metric being constructed from
the latter by the relation
\[
g^{ab} = - \frac{1}{\dv}
\Tr ({\gamma}^a {\gamma}^b).
\]
The variation of the connection $\mathbf{A}$
tells us that
\begin{equation}
D_b \left( g^{1/2} {\gamma}^{ab} \right) = 0.
\label{pal1}
\end{equation}

This equation, by itself, is not enough to uniquely determine
the connection. This, however, is not a problem unique to our
spinorial approach. If one considers the standard Palatini approach to
the Einstein-Hilbert action, then the equations which follow
from variation of the connection is supposed to uniquely determine
the connection as the Levi-Civita connection. On closer inspection,
however, this turns out not to be the case. If one, a priori, assumes
the connection is torsion-free, then the equations of motion tell us
that the connection is metric, and vice versa. However, if we start
with a completely general connection, the equations of motion in the
Palatini formalism are insufficient to uniquely determine the
connection.

It is therefore important to consider what we would like to demand of
a connection, and what further conditions we must impose, by hand, on
the connection so that the equations of motion uniquely confine it to be
the spin-connection.

The complex Clifford algebra
$\mathbb{C}\mathrm{l}_n$
has a unique irreducible representation
on $\mathbb{C}^{\dv}$ when $n$ is even, and two
inequivalent irreducible representations on
$\mathbb{C}^{\dv}$ when $n$ is odd, where
\[
{\dv} =
\begin{cases}
2^{n/2}& \mbox{$n$ even}
\\
2^{(n-1)/2}& \mbox{$n$ odd}.
\end{cases}
\]
Therefore, assume that have an irreducible representation of our
Clifford algebra on $\mathbb{V} \cong \mathbb{C}^{\dv}$. The
elements of the Clifford algebra are then represented as
endomorphisms of $\mathbb{V}$ so, relative to any basis for
$\mathbb{V}$, would correspond to elements of $\mathbb{C}(\dv)$,
the set of ${\dv} \times {\dv}$ complex matrices.

Given an irreducible representation of the algebra on a space
$\mathbb{V}$, we may construct bi-linear forms
\begin{align*}
{}^{\pm}{\epsilon}: &\mathbb{V} \otimes \mathbb{V} \rightarrow
\mathbb{C},
&
{}^{\pm}{\epsilon}^*: &\mathbb{V}^* \otimes \mathbb{V}^* \rightarrow
\mathbb{C}
&
&\mbox{$n$ even},
\\
{\epsilon}: &\mathbb{V} \otimes \mathbb{V} \rightarrow
\mathbb{C},
&
{\epsilon}^*: &\mathbb{V}^* \otimes \mathbb{V}^* \rightarrow
\mathbb{C}
&
&\mbox{$n$ odd},
\end{align*}
with the symmetry
properties shown in Table~\ref{epsymm} \cite{PR}.
(In odd dimensions, $\epsilon$ shall denote the
one of ${}^\pm \epsilon$ which is non-vanishing.)

\begin{table}[ht!]
\begin{center}
\begin{tabular}{|>{$}l<{$}|>{$}l<{$}|>{$}l<{$}|}
\hline
\multicolumn{1}{|c|}{$n \pmod 8$}
& \multicolumn{1}{c|}{${}^{+}\epsilon$} &
\multicolumn{1}{c|}{${}^{-}\epsilon$}\\
\hline\hline
0 & \mbox{Symmetric} & \mbox{Symmetric} \\
1 & \mbox{zero} & \mbox{Symmetric} \\
2 & \mbox{Skew-symmetric} & \mbox{Symmetric} \\
3 & \mbox{Skew-symmetric} & \mbox{zero} \\
4 & \mbox{Skew-symmetric} & \mbox{Skew-symmetric} \\
5 & \mbox{zero} & \mbox{Skew-symmetric} \\
6 & \mbox{Symmetric} & \mbox{Skew-symmetric} \\
7 & \mbox{Symmetric} & \mbox{zero} \\
\hline
\end{tabular}
\end{center}
\vspace{8pt}
\caption{Symmetries of ${}^{\pm}\epsilon$
in various dimensions}
\label{epsymm}
\end{table}

One can show that the forms ${}^\pm \epsilon$ have
the properties that, for $p = 1, \dots n$
\begin{equation}
{}^{\pm} \epsilon ( \lambda, \gamma^{A_1 \dots A_p} \phi ) =
(\mp 1)^p \, (-1)^{\frac{p(p-1)}{2}} \,
{}^{\pm} \epsilon ( \gamma^{A_1 \dots A_p} \lambda, \phi ), \qquad
\forall \lambda, \phi \in \mathbb{V}.
\label{gammsymm}
\end{equation}

If we now consider a spin manifold $X$, with metric
$\mathbf{g}$, then all of the above algebraic
considerations carry across to the Clifford
algebra bundle over $X$. This is the bundle
generated by Clifford multiplication from
the cotangent bundle $T^* X$, and the
representation space $\mathbb{V}$ becomes the spin
bundle, the sections of which are spinor fields.
Since the maps ${}^{\pm} \epsilon$ are suitably
equivariant under $\mathrm{Spin}_{r, s}$ transformations,
they carry across directly to corresponding forms on
the spin-bundles.

Given an orthonormal frame for $T^* X$,
there is a natural connection on $T^* X$,
the spin-connection. One can lift this connection to a unique
connection on the spin-bundle $\mathbb{V}$.
We wish to consider a minimal set of spinorial
conditions we can impose on a connection on $\mathbb{V}$ which
will uniquely define it to be this image of the spin-connection.
Given a connection on $T^* X$, we can extend this to a connection
on $\Lambda^* X$. One would then like to
define a connection on the Clifford
algebra bundle which commutes
with this map $\boldsymbol{\sigma}$
introduced above
\begin{equation}
\left( \nabla_{\mathbf{X}} \circ \boldsymbol{\sigma} \right)
\boldsymbol{\lambda} =
\boldsymbol{\sigma}
\left( \nabla_{\mathbf{X}} \boldsymbol{\lambda} \right).
\label{gammascovconst}
\end{equation}
In colloquial terms, this means that if we view the
collection of $\gamma$-matrices as a section of $TX \otimes
\mathrm{End} (\mathbb{V}) \cong TX \otimes \mathbb{V}^* \otimes
\mathbb{V}$ then, given the connection on $TX$, we wish to
arrange the connection on $\mathbb{V}$ so that this section is
covariantly constant. This does not uniquely
determine the connection. However, if we impose
the additional requirement on the connection on
$\mathbb{V}$ that it annihilates the forms ${}^{\pm} \epsilon$,
this uniquely determines the connection on $\mathbb{V}$
to be the image of the spin-connection defined
above.

What we need to know, however, is the minimal set of conditions we
must impose on a connection on $\mathbb{V}$ in order that when
combined with the equation of motion~\eqref{pal1} the connection is
uniquely determined to be the image of the spin connection. One
requirement would be that the connection should annihilate the
${}^{\pm} \epsilon$
\begin{align}
\nabla {}^{\pm} \epsilon &= 0 & &\mbox{$n$ even},
\nonumber\\
\nabla \epsilon &= 0 & &\mbox{$n$ odd}.
\label{presep}
\end{align}
The question is to what extent this condition
determines the connection. If we consider
two connections $\nabla$ and ${\nabla}^{\prime}$ on $\mathbb{V}$,
then
\[
{\nabla}^{\prime}_{\mathbf{X}} \psi -
\nabla_{\mathbf{X}} \psi =
< \mathbf{T}, \mathbf{X} > \psi, \qquad
\forall \mathbf{X} \in \Gamma (TX), \qquad
\forall \psi \in \Gamma (\mathbb{V}),
\]
where $\mathbf{T}$ is a section of $\Lambda^1 (X) \otimes
\mathrm{End}(\mathbb{V})$ that transforms under the adjoint
representation under the $\mathrm{Spin}_{r, s}$ action on $\mathbb{V}$.
If we assume that both connections annihilate the forms
${}^{\pm} \epsilon$, then we find that we require
\[
{}^{\pm} \epsilon ( \lambda, < \mathbf{T}, \mathbf{X}> \phi ) +
{}^{\pm} \epsilon ( < \mathbf{T}, \mathbf{X}> \lambda, \phi ) = 0, \qquad
\forall \lambda, \phi \in \Gamma (\mathbb{V}),
\qquad \forall \mathbf{X} \in \Gamma (TX).
\]
{}From Equation~\eqref{gammsymm}, we therefore deduce that
$< \mathbf{T}, \mathbf{X}>$ is an section of
$\mathrm{Im}_{\sigma} \left( \Lambda^2 \oplus \Lambda^6 \oplus \dots
\right) \subset \mathbb{C}\mathrm{l}_{r, s}$ for all vector fields $\mathbf{X}$.
Therefore
\begin{equation}
\mathbf{T} \in \Gamma \left(
\Lambda^1 (X) \otimes
\mathrm{Im}_{\sigma} \left( \Lambda^2 \oplus \Lambda^6 \oplus \dots
\right) \right).
\label{texp}
\end{equation}
(The expansion on the right-hand-side of this equation terminates when
we reach the highest integer $4k+2$ less than or equal to $n$.) We
should perhaps note that the forms ${}^\pm \epsilon$ are not uniquely
determined by the Clifford algebra, but only determined up to a scale.
Therefore a choice of this scale is implicit in Equation~\eqref{presep}.
One could equivalently work without fixing this scale, imposing only the
existence of $1$-forms ${}^\pm \boldsymbol{\lambda}$ with the property
that $\nabla {}^{\pm} \epsilon = {}^{\pm} \boldsymbol{\lambda} \otimes
{}^{\pm} \epsilon$ on the connection. If we then impose the additional
condition that connection is trace-free (i.e. is an
$\mathrm{SL}(\mathbb{V})$ connection), then we recover the
result~\eqref{texp}.

We now wish to impose the further condition on our connection that it
satisfies Equation~\eqref{pal1}, which followed from our action
principle considerations. If we assume that the connections $\nabla$ and
$\nabla^{\prime}$ obey this equation, then the field $\mathbf{T}$
obeys the condition
\begin{equation}
\left[ T_a, \gamma^{ab} \right] = 0.
\label{teom}
\end{equation}
Since pure elements of order $2$ in the Clifford algebra generate
$\mathrm{Spin}_{r, s}$ transformations, Equation~\eqref{teom} must be
separately satisfied by each of the components in the Clifford algebraic
expansion of $\mathbf{T}$ given in Equation~\eqref{texp}.
First consider dimensions $n
\neq 2 \pmod 4$. Denote the part of $\mathbf{T}$ of order $m=4k+2$ in
the Clifford algebra by $\mathbf{T}_m$. This can be viewed as a section
of $\wedge^1 (X) \otimes \wedge^m (X) \cong \wedge^{m+1}(X) \oplus
\wedge^{m-1}(X) \oplus S_0^{m+1}(X)$, where the separate terms
correspond to the totally skew-symmetric, trace, and trace-free parts of
$\mathbf{T}_m$, respectively. We then use the simple result:

\begin{lemma}
\label{lem2}
\[
\left[
\gamma_{AB},
\gamma^{C_1 C_2 \dots C_p}
\right] =
4 \, p \, \delta_{[A}^{[C_1}
\gamma_{B]}{}^{C_2 \dots C_p]},
\qquad
p = 1, \dots n.
\]
\end{lemma}

The proof of this result is a straightforward application of
definitions, and so will be omitted. With this result, it follows
from the relation~\eqref{teom} that the $\wedge^{m+1}(X)$ and
$\wedge^{m-1}(X)$ parts of $\mathbf{T}_m$ vanish. The remaining
equations are then inconsistent with the symmetries required of an
element of $S_0^m$ (this is a generalisation of the argument that
any $(0,3)$ tensor, $\mathbf{a}$, with the symmetry property
$\mathbf{a} (\mathbf{x}, \mathbf{y}, \mathbf{z}) = \mathbf{a}
(\mathbf{y}, \mathbf{x}, \mathbf{z}) = - \mathbf{a} (\mathbf{x},
\mathbf{z}, \mathbf{y})$ must vanish identically), and so the only
solution of our requirements is $\mathbf{T}_m = 0$, and therefore
$\mathbf{T} = 0$. Therefore, if $n \neq 2 \pmod 4$, the connection
which annihilates the forms ${}^{\pm} \epsilon$ and which satisfies
the equations of motion~\eqref{pal1} is unique, and must be the
pull-back to the spin-bundle of the spin-connection, as required.

In dimension $n=4k+2$, however, this argument breaks
down. If we consider the final term in the expansion~\eqref{texp},
we have the possibility
\[
\mathbf{T} = \boldsymbol{\phi} \otimes \omega
\]
for any $1$-form field $\boldsymbol{\phi}$, where
\begin{equation}
\omega = i^{\alpha}
{\gamma}^1 \cdots {\gamma}^n,
\qquad
\alpha = \left[ \frac{n+1}{2} \right]
\label{volelement}
\end{equation}
is the volume element on the Clifford algebra \cite{LM}.
With respect to a general connection of this form, the
$\gamma$-matrices are not covariantly constant, although a
skew-symmetrised product of an even number of $\gamma$-matrices is
covariantly constant. Since such a term has no analogue in a
tensorial approach, it does not seem to have any straightforward
geometrical interpretation.

If we wish to reproduce Einstein-Hilbert gravity
it is therefore necessary, in these dimensions, to impose
an additional condition on the spinor connection to remove this extra
freedom. We know,
from above, that
any connection which preserves the forms
${}^{\pm} \epsilon$ must be in the image
of $\Lambda^2 \oplus \Lambda^6 \oplus
\dots \oplus \Lambda^n$. The most direct
way to remove the extra freedom of equation
\eqref{extra} is therefore to impose by
hand the condition that there is no
$\Lambda^n$ term:
\begin{equation}
\mathbf{A} \in \Gamma \left( \Lambda^1 (X) \otimes
\mathrm{Im}_{\sigma} \left( \Lambda^0 (X) \oplus
\Lambda^1 (X) \oplus \dots \oplus
\Lambda^{n-1} (X) \right) \right) \qquad
n \equiv 2 \pmod 4.
\label{extra}
\end{equation}
This is equivalent to imposing the condition
\[
\Tr \left( \mathbf{A} \omega \right) = 0
\]
on the connection.
Only once this extra degree of freedom has
been removed from the theory do we find
that the connection defined by
Equations~\eqref{texp} and
\eqref{teom} is the connection we require.

Note that, in principle, the algebraic conditions~\eqref{presep} and
\eqref{extra} on the connection could be imposed as extra primary
constraints in the Lagrangian approach we have
taken. Since these constraints have no dynamics, however, the resulting
theory will be identical with the one which results by simply assuming
that the connection obeys these constraints identically. For simplicity,
we shall adopt the latter approach.

In even dimensions, we can use $\omega$ to split up $\mathbb{V}$
into chiral parts $\mathbb{V}^{\pm}$. A significant difference
between dimensions $0
\pmod 4$ or $2
\pmod 4$ is that we can arrange that
\begin{align*}
{}^+ \epsilon:
\mathbb{V}^+ \otimes \mathbb{V}^+
&\rightarrow \mathbb{C},
&
{}^- \epsilon:
\mathbb{V}^- \otimes \mathbb{V}^-
&\rightarrow \mathbb{C},
&
n &\equiv 0 \pmod 4,
\\
{}^+ \epsilon:
\mathbb{V}^+ \otimes \mathbb{V}^-
&\rightarrow \mathbb{C},
&
{}^- \epsilon:
\mathbb{V}^- \otimes \mathbb{V}^+
&\rightarrow \mathbb{C},
&
n &\equiv 2 \pmod 4.
\end{align*}
Therefore ${}^{\pm} \epsilon$ define isomorphisms
\begin{align*}
\mathbb{V}^{\pm} &\cong (\mathbb{V}^{\pm})^*,
&
n &\equiv 0 \pmod 4,
\\
\mathbb{V}^{\pm} &\cong (\mathbb{V}^{\mp})^*,
&
n &\equiv 2 \pmod 4.
\end{align*}
The fact that these isomorphisms are between different spin-bundles
in dimensions $2 \pmod 4$ means that in these dimensions we cannot
reduce the problem to a single spin-bundle, and a chiral spinorial
formalism does not seem to be possible. In dimensions $0 \pmod 4$,
however, it is possible to set up the theory completely on, for
example, the positive spin-bundle $\mathbb{V}^+$. In odd
dimensions, the volume element is a central element of the Clifford
algebra, so if we consider irreducible spin-bundles, there is no
similar reduction.

Returning to the equations of motion, if we assume that we have
imposed these extra conditions in such a way that the connection
$\mathbf{A}$ is identified with the spin connection, and therefore
its curvature is identified with the curvature of the
spin-connection as in Equation~\eqref{FR}, then varying the
$\gamma$-matrices in the action~\eqref{ehaction}, we find,
\[
\delta S_{EH} = \frac{1}{2} {\int}_X g^{1/2}
\Tr \left( (\delta {\gamma}^a) {\gamma}_b
(r_a{}^b - \frac{1}{2} s {\delta}_a{}^b )\right)
\, d^n x.
\]
Therefore the equation of motion tells us that
the metric $\mathbf{g}$ satisfies the vacuum Einstein
equations
\[
\mathbf{r} = \frac{s}{2} \mathbf{g}.
\]

As such, as long as we impose the condition that the connection
$\mathbf{A}$ annihilates the natural bi-linear forms on the spin
bundle, along with the extra condition of no volume terms in
dimensions $2 \pmod 4$, then the equations of motion which follow
from our version of the Einstein-Hilbert action are equivalent to
the vacuum Einstein equations.

\section{Riemannian Hamiltonian Decomposition}
\label{sec:hamdec}

We now wish to consider the Hamiltonian
version of the above theory. For simplicity,
we will consider metrics of Riemannian signature,
although other signatures can be treated similarly.
Our treatment initially follows the standard
approach in four dimensions \cite{JS,S1}.

We consider a suitable open set $U \subset X$
of our manifold which we assume to be foliated
by a $1$-parameter family of leaves $\Sigma$ of
dimension $(n-1)$. Introducing a parameter $t$ to
label the different leaves of this foliation, and local coordinates $\{
x^i | i = 1, \dots , n-1 \}$ on $\Sigma$, we may decompose the metric
in standard Hamiltonian form
\[
g = \boldsymbol{\epsilon}^0
\otimes \boldsymbol{\epsilon}^0
+ {\delta}_{IJ} \boldsymbol{\epsilon}^I
\otimes \boldsymbol{\epsilon}^J,
\]
where we take
\begin{eqnarray*}
\boldsymbol{\epsilon}^0 &=& N dt,
\\
\boldsymbol{\epsilon}^I &=& e_i{}^I
\left( dx^i + N^i dt \right).
\end{eqnarray*}
(Upper case indices $I, J, \dots$ take values $1, 2, \dots, n-1$.)
The dual basis takes the form
\begin{eqnarray*}
\mathbf{e}_0 &=& N^{-1}
\left( {\partial}_t
- N^i {\partial}_i \right),
\\
\mathbf{e}_I &=& e_I{}^i {\partial}_i.
\end{eqnarray*}
The induced metric (first fundamental form) on $\Sigma$ will be
denoted $\mathbf{q}$, and has components
\[
q_{ij} = {\delta}_{IJ} e_i{}^I e_j{}^J
\]
with respect to the coordinates introduced above.

At this point it is
useful to recall the
Clifford algebra isomorphism
\[
\mathbb{C}\mathrm{l}_{n}^{\mathrm{even}} \cong
\mathbb{C}\mathrm{l}_{n-1}.
\]
In our context, this means that we may define the
$(n-1)$-dimensional $\gamma$-matrices by
\[
\Gamma^I := \gamma^{0I}, \qquad
I = 1, \dots , n-1,
\]
which generate the algebra $\mathbb{C}\mathrm{l}_{n-1}$. We also
define the spatial $\gamma$-matrices
\[
\Gamma^i := \Gamma^I \, e_I^i,
\]
and their skew-symmetrised products
$\Gamma^{ij\dots k}$.

If we now insert the decomposition of
the metric into the Einstein-Hilbert
action, it takes the form
\[
S = {\int}_X q^{1/2} \Tr
\left[{\Gamma}^I
< \mathcal{L}_t \boldsymbol{\gamma}, \mathbf{e}_I> -
{\Gamma}^i D_i^{\gamma} A_t
+ {\Gamma}^i F_{ij}^{\gamma} N^j
+ \frac{N}{2} F_{ij}^{\gamma}
{\Gamma}^{ij} \right] \, dt \, d^{n-1}x.
\]
In this equation we have defined the connection
\[
\boldsymbol{\gamma} = p_{\Sigma} ( \mathbf{A} )
\]
as the pull-back of the connection $\mathbf{A}$ to the surface
$\Sigma$. The curvature of this connection, $\boldsymbol{\gamma}$
is denoted $\mathbf{F}^{\gamma}$, and we have defined the covariant
derivative of the field $A_t$ by
\[
D_i A_t = {\partial}_i A_t +
\left[ {\gamma}_i, A_t \right].
\]
We also, for simplification later,
introduce a densitised version of the
function $N$ by defining
\[
{\underline N} := q^{-1/2} N,
\]
where
\[
q := | \mathrm{det} \left( q_{ij} \right) |.
\]

In order to proceed with the
Hamiltonian decomposition, we introduce
momenta conjugate to all of the
dynamical variables. {}from the form of
the Lagrangian above, we deduce
that the momenta conjugate to the
variables $({\gamma}_i, e_I{}^i,
A_t, {\underline N}, N^i)$, in
order, take the form:
\begin{eqnarray*}
{\pi}^i &=& {\tilde \sigma}^i,
\\
{\pi}^I{}_i &=& 0,
\\
{\pi}^t &=& 0,
\\
{\underline \pi} &=& 0,
\\
{\pi}_i &=& 0,
\end{eqnarray*}
where we have defined the densitised
$\gamma$-matrices
\[
{\tilde \sigma}^i := q^{1/2} \Gamma^i.
\]
In Dirac's terminology \cite{D}, we therefore have the primary
constraints of the theory
\begin{subequations}
\label{primcons}
\begin{align}
{\phi}^i &= {\pi}^i - {\tilde \sigma}^i,
\label{pc1}
\\
{\phi}^I{}_i &= {\pi}^I{}_i,
\label{pc2}
\\
{\phi}^t &= {\pi}^t,
\label{pc3}
\\
{\underline \phi} &= {\underline \pi},
\label{pc4}
\\
{\phi}_i &= {\pi}_i.
\label{pc5}
\end{align}
\end{subequations}
The total Hamiltonian $H_T$ of the theory
is now the sum of the
canonical Hamiltonian $H_c \sim
p {\dot q} - L$ and primary constraints with
suitable Lagrange multipliers
\begin{multline*}
H_T = - \int_{\Sigma}
\Tr
\left[
A_t D_i^{\gamma} {\tilde\sigma}^i +
F_{ij}^{\gamma} {\tilde\sigma}^i N^j +
\frac{1}{2} {\underline N}
F_{ij}^{\gamma} {\tilde\sigma}^i {\tilde\sigma}^j
\right]
\\
+ \int_{\Sigma}
\Tr \left( {\lambda}_i {\phi}^i \right)
+ {\lambda}_I{}^i {\phi}^I{}_i
+ \Tr \left( {\lambda}_t {\phi}^t \right)
+ {\underline \lambda} {\underline \phi}
+ {\lambda}^i {\phi}_i.
\end{multline*}
Time evolution is generated by Poisson Brackets with this Hamiltonian,
with $\mathcal{L}_t f = \left[ f, H_T \right]$ with $f$ any function on
the phase space, and where the variables $q = ({\gamma}_i, e_I{}^i,
A_t, {\underline N}, N^i)$ and conjugate momenta
$p = ({\pi}^i, {\pi}^I{}_i, {\pi}^t,
{\underline \pi}, {\pi}_i)$ obey the
heuristic Poisson Bracket relations
\begin{equation}
\left[ q, p \right] = 1.
\label{heurpbs}
\end{equation}

\subsection{Secondary Constraints}
\label{sec:sec}

The primary constraints of the theory~\eqref{primcons} must be
preserved under time evolution. In order that the constraints
${\phi}^t, {\phi}_i, {\underline \phi}$ be preserved by evolution,
their Poisson Bracket with $H_T$ must be a sum of constraints. This
leads to the secondary constraints of the theory
\begin{subequations}
\label{primaryconstraints}
\begin{align}
{\chi}_1 &:= D_i^{\gamma} {\tilde\sigma}^i \approx 0,
\\
{\chi}_{2i} &:= - \Tr
\left( F_{ij}^{\gamma} {\tilde\sigma}^j \right) \approx 0,
\\
{\chi}_3 &:= \frac{1}{2}
\Tr \left( F_{ij}^{\gamma}
{\tilde\sigma}^i
{\tilde\sigma}^j
\right) \approx 0.
\end{align}
\end{subequations}
The preservation of the constraints
$\chi_1, \chi_{2i}, \chi_3$ do not
lead to any new secondary constraints
of the theory, but simply place
restrictions on the Lagrange multipliers
\begin{align*}
\mathcal{L}_t \chi_1 &\approx
\left[ \lambda_i, {\tilde\sigma}^i \right] \approx 0,
\\
\mathcal{L}_t \chi_{2i} &\approx - \Tr
\left[ \left( D_i^{\gamma} \lambda_j -
D_j^{\gamma} \lambda_i \right) {\tilde\sigma}^j \right]
\approx 0,
\\
\mathcal{L}_t \chi_3 &\approx \frac{1}{2} \Tr
\left[ D_i^{\gamma} \lambda_j
\left[ {\tilde\sigma}^i, {\tilde\sigma}^j \right]\right]
\approx 0.
\end{align*}
If we consider the equations of
motion for the corresponding variables
$( A_t, N^i, {\underline N} )$,
we find
\[
\mathcal{L}_t A_t = {\lambda}_t, \qquad
\mathcal{L}_t {\underline N} =
{\underline \lambda}, \qquad
\mathcal{L}_t N^i = {\lambda}^i.
\]
Since the constraints ${\phi}^t,
{\phi}_i, {\underline \phi}$
have vanishing Poisson Brackets
with all the other constraints,
we may drop the momenta ${\pi}^t,
{\pi}_i, {\underline \pi}$,
and the multipliers ${\lambda}_t,
{\lambda}^i, {\underline \lambda}$,
and simply view $A_t, N^i,
{\underline N}$ as Lagrange multipliers
enforcing the constraints
\eqref{primaryconstraints}.

Eliminating these redundant
variables and constraints implies that
we are left with the dynamical
variables of the theory
\[
({\gamma}_i, \pi^i; e_I{}^i, \pi^I{}_i),
\]
and the constraints
\begin{subequations}
\label{fincon}
\begin{align}
{\phi}^i &= {\pi}^i - {\tilde\sigma}^i,
\\
{\phi}^I{}_i &= {\pi}^I{}_i,
\\
{\chi}_1 &= D_i^{\gamma} {\tilde\sigma}^i,
\\
{\chi}_{2i} &= - \Tr
\left( F_{ij}^{\gamma} {\tilde\sigma}^j \right),
\\
{\chi}_3 &= \frac{1}{2}
\Tr \left( F_{ij}^{\gamma} {\tilde\sigma}^i
{\tilde\sigma}^j \right).
\end{align}
\end{subequations}
The Hamiltonian of the theory
is simply the sum of these constraints
multiplied by the appropriate
Lagrange multipliers
\[
H_T = - \int_{\Sigma} \left( \Tr
\left[ A_t \chi_1 \right]
+ N^i \chi_{2i}
+ {\underline N} \chi_3 \right)
+ \int_{\Sigma}
\left( \Tr \left( {\lambda}_i {\phi}^i \right)
+ {\lambda}_I{}^i {\phi}^I{}_i \right).
\]
Since the Hamiltonian is a sum
of constraints, the preservation of these
constraints under time evolution
reduces to a problem concerning the
Poisson Brackets of the constraints.
If a constraint is first-class,
then it will automatically be
preserved by the evolution, whereas if it
is second-class, its preservation
will place restrictions on the
Lagrange multipliers. In neither
case will preservation under time
evolution introduce new constraints,
so the only constraints of the
theory are those given in \eqref{fincon}.

\subsection{Removal of Higher Order Constraints}
\label{sec:higher}

We know from the arguments of
Section~\ref{sec:eom} that the connection
$\mathbf{A}$ must obey the extra
geometrical constraint that it
annihilates the forms
${}^{\pm} \epsilon$ on the
spin-bundle. In terms
of the Hamiltonian theory,
this implies that the spinorial quantities
$(A_i, \pi^i, A_t)$ take values
in $\mathrm{Im}_{\sigma} (\Lambda^2 (X) \oplus
\Lambda^6 (X) \oplus \dots) \subset
\mathbb{C}\mathrm{l}_n$, where the expansion
terminates at $\Lambda^{4k+2}(X)$ for the
largest value of $k$ with $4k+2 \le n$. This, in turn,
implies that the constraints
$(\phi^i, \chi_1)$ take values in the same
space. As such, we see that new constraints
enter the theory in each dimension $2 \pmod 4$.
In line with the arguments of Section~\ref{sec:eom}, however, we know
that in these dimensions we must impose the extra condition that
the Clifford algebra expansion of the
spinor-connection (and therefore the conjugate momentum) must not
contain any multiple of the volume form in order to recover standard
Einstein-Hilbert gravity. Therefore, we must also, by hand, remove the
corresponding constraints which arise. As such, the new constraints really
only come into effect in dimensions $3 \pmod 4$.

It now becomes useful to
divide the constraints $(\phi^i,
\chi_1)$ and the variables
$(A_i, \pi^i, A_t)$ into a pure second order
part taking values in
$\mathrm{Im}_{\sigma} \,
\Lambda^2 (X) \subset \mathbb{C}\mathrm{l}_n$
and a part taking values in the
higher order space $\mathrm{Im}_{\sigma}
(\Lambda^6 (X) \oplus \Lambda^{10} (X)
\oplus \dots) \subset \mathbb{C}\mathrm{l}_n$.
Further decomposing using the
isomorphism $\mathbb{C}\mathrm{l}_n^{\mathrm{even}}
\cong \mathbb{C}\mathrm{l}_{n-1}$
introduced in Section~\ref{sec:hamdec}, we define
\begin{align*}
\gamma_i &:= A_i + {\tilde A}_i,
\\
\pi^i &:= \Pi^i + {\tilde \Pi}^i,
\\
\phi^i &:= \Phi^i + {\tilde \Phi}^i,
\\
\chi_1 &:= \Psi + {\tilde \Psi},
\end{align*}
where
\[
A_i, \Pi^i, \Phi^i, \Psi
\in
\mathrm{Im}_{\sigma} (\Lambda^1 (\Sigma)
\oplus \Lambda^2 (\Sigma)),
\qquad
{\tilde A}_i, {\tilde \Pi}^i, {\tilde \Phi}^i, {\tilde \Psi}
\in
\mathrm{Im}_{\sigma} (\Lambda^5 (\Sigma)
\oplus \Lambda^6 (\Sigma) \oplus \dots).
\]
We can therefore write
\begin{align*}
\Phi^i &= \Pi^i - {\tilde\sigma}^i,
&{\tilde\Phi}^i &= {\tilde\Pi}^i,
\\
\Psi &= \partial_i {\tilde\sigma}^i +
\left[ A_i, {\tilde\sigma}^i \right],
&{\tilde\Psi} &=
\left[ {\tilde A}_i, {\tilde\sigma}^i \right].
\end{align*}
Generally, we will refer to the quantities of higher order in the
Clifford algebra i.e. ${\tilde A}_i,
{\tilde \Pi}^i, {\tilde \Phi}^i, {\tilde \Psi}$ as
\lq\lq higher order\rq\rq\ quantities. Our first goal
is to show that the higher order variables and constraints
may be removed from the theory, leaving a theory with only
the first and second order quantities in the Clifford algebra,
i.e. $A_i, {\Pi}^i, {\Phi}^i, \Psi$. To this end, we
make use of the following straightforward result:

\begin{lemma}
\label{lem1}
The trace of the skew-symmetrised $\gamma$-matrices
\[
\Tr \left( \gamma^{A_1 \dots A_{2l}} \gamma_{B_1 \dots B_{2m}}
\right)
\]
vanishes unless $l = m$ and the indices $A_1, \dots A_{2l}$ are a
permutation of the indices $B_1 \dots B_{2m}$. In particular,
\[
\Tr \left( \gamma^{A_1 \dots A_k} \gamma_{B_1 \dots B_k}
\right) = (-1)^{\frac{k(k+1)}{2}} \, {\dv} \, k! \,
\delta^{[A_1}_{B_1} \cdots \delta^{A_k]}_{B_k},
\]
where $\dv$ is the rank of the spin-bundle.
\end{lemma}

Assuming linear independence of the
vector fields $\mathbf{e}_I$ on $\Sigma$,
then $\mathbb{C}\mathrm{l}_{n-1}$ will be spanned by the elements
${\tilde\sigma}^i$ and their skew-symmetrised products.
We may therefore decompose each of the
higher order spinorial quantities above
into pure constituent parts in the
Clifford algebra using the ${\tilde\sigma}^i$.
We define
\begin{align*}
\phi^{ij_1 \dots j_m} &:=
\Tr \left[
{\tilde\Phi}^i {\tilde\sigma}^{[j_1} \dots
{\tilde\sigma}^{j_m]} \right],
\\
\psi^{j_1 \dots j_m} &:=
\Tr \left[
{\tilde\Psi}
{\tilde\sigma}^{[j_1} \dots {\tilde\sigma}^{j_m]}
\right].
\end{align*}
where $m = 4k+1$ or $m = 4k+2$
with $k = 1, 2, \dots$.

These constraints are not all first-class,
and our first task is to identify the
second-class constraints and remove them
by Dirac's procedure.
In order to do this,
we consider the constraints
$\phi^{ij_1 \dots j_m}$ in terms of SO$_{n-1}$
representation theory. We can represent these
constraints as an element of
$\wedge^1 (\Sigma) \otimes \wedge^m
(\Sigma)$ and use the SO$_{n-1}$
decomposition
\begin{equation}
\wedge^1 (\Sigma) \otimes \wedge^m (\Sigma)
\cong
\wedge^{m+1} (\Sigma) \oplus
S_0^{m+1} (\Sigma) \oplus
\wedge^{m-1} (\Sigma).
\label{sondecomp}
\end{equation}
In this decomposition,
$\phi \in \wedge^1 (\Sigma) \otimes \wedge^m
(\Sigma)$ decomposes into a totally skew-symmetric
part, denoted $\wedge^{m+1} [\phi]$,
and a part of mixed symmetry,
$S^{m+1} [\phi]$. This second
part may then be reduced to irreducible
components consisting of a trace-free
element $S_0^{m+1} [\phi] \in S_0^{m+1} (\Sigma)$, and a
trace part, which may then be identified
with an element of $\wedge^{m-1} (\Sigma)$,
denoted $\wedge^{m-1} [\Tr \phi]$.

Similarly, ${\tilde \Psi} \in
\mathrm{Im}_{\sigma} (\Lambda^5 (\Sigma)
\oplus \Lambda^6 (\Sigma)
\oplus \dots)$ and we denote the separate
elements of this decomposition by
$\wedge^{m} [\psi], m = 4k+1,
4k+2$.

Recall that Dirac's procedure for eliminating
second-class constraints from finite dimensional
systems consists of finding
a set of second class constraints
$\{ \alpha_i | i = 1, \dots , k \}$ with the
property that the matrix of Poisson Brackets
$C_{ij} := \left[ \alpha_i , \alpha_j \right]$
is of maximal rank \cite{D}. One then defines the
Dirac bracket of two arbitrary functions $f$
and $g$ on the phase space by the modified relation
\begin{equation}
\left[ f, g \right]^* :=
\left[ f, g \right] -
\left[ f, \alpha_i \right]
\left( C^{-1} \right)^{ij}
\left[ \alpha_j, g \right].
\label{dirbrac}
\end{equation}
The Dirac bracket automatically
has the property that
$\left[ f, \alpha_i \right]^* = 0$
for any function $f$ on the phase space. Similarly,
given a function $f$ on the phase space, we
define a new function,
\begin{equation}
f^* :=
f -
\left[ f, \alpha_i \right]
\left( C^{-1} \right)^{ij}
\alpha_j.
\label{dirfn}
\end{equation}
The constraints automatically vanish, in the sense that
$\alpha_i^* = 0$. The assumption of the Dirac approach is
that the dynamics of the theory, with second class constraints
removed, is generated by a Hamiltonian $H^*$ via Dirac Brackets
\[
\frac{d}{dt} f^* := \left[ f^*, H^* \right]^*.
\]
In practice, this means we remove the
redundant variables of the theory by imposing the
constraints as identities, and then define the modified Poisson Brackets
of the remaining quantities to be the Dirac Bracket.

The procedure generalises to field
theories with constraints in the obvious fashion
(although a choice of boundary conditions is
generally involved in constructing the inverse
matrix $C^{-1}$).
In our case, the goal is therefore to find pairs of
families of second-class
constraints which are canonically conjugate, in the sense that
the matrix of Poisson Brackets of the constraints is of maximal
rank. Using the Poisson Brackets
\eqref{heurpbs}, we find that two such families of
conjugate second-class constraints are
given by
\begin{align}
\wedge^{4k+1} [\psi]
&\leftrightarrow \wedge^{4k+1} [\Tr \phi],
\label{high1}
\\
\wedge^{4k+2} [\psi]
&\leftrightarrow \wedge^{4k+2} [\phi].
\label{high2}
\end{align}
These constraints have vanishing Poisson Brackets with
all other variables and constraints of the theory,
and they simply constrain the variables
$( \wedge^{4k+1} [{\tilde A}],
\wedge^{4k+2} [{\tilde A}])$ and their
conjugate momenta
$(\wedge^{4k+1} [\Tr {\tilde\Pi}],
\wedge^{4k+2}[{\tilde\Pi}])$ to vanish.
Therefore, when we construct the Dirac brackets as in
Eq.~\eqref{dirbrac}, these constraints have vanishing Dirac bracket
with all other functions on the phase space. All other Dirac Brackets
remain equal to the original Poisson Brackets. Similarly, when
we redefine functions on the phase
space as in Eq.~\eqref{dirfn}, this sets the variables
$( \wedge^{4k+1} [{\tilde A}],
\wedge^{4k+2} [{\tilde A}])$ and their
conjugate momenta
$(\wedge^{4k+1} [\Tr {\tilde\Pi}],
\wedge^{4k+2}[{\tilde\Pi}])$ identically to zero,
whilst all other functions
on the phase space remain unchanged. Therefore
the constraints~\eqref{high1} and \eqref{high2}
simply set the variables
$( \wedge^{4k+1} [{\tilde A}],
\wedge^{4k+2} [{\tilde A}])$ and momenta
$(\wedge^{4k+1} [\Tr {\tilde\Pi}],
\wedge^{4k+2}[{\tilde\Pi}])$ identically to zero,
so that these variables and momenta are removed
completely from the theory.

Carrying out this procedure for each value of $k \ge 1$
completely removes all of the higher order constraints ${\tilde \Psi}$
from the theory, leaving us with only the first and
second order part of $\chi_1$, denoted $\Psi$ above.
The remaining higher order parts of the constraint
$\phi^i$ impose that the remaining
parts of the momenta ${\tilde \Pi}$
vanish. Preservation of these constraints
under time evolution then places restrictions on
the Lagrange multipliers of the theory. These
restrictions, modulo the second-class constraints
discussed above, when combined with
the explicit form of the metric imply that
\[
\left[ {\tilde A}_a, \gamma^{ab} \right] = 0.
\]
It is straightforward to show, in dimensions
$n \neq 2 \pmod 4$, that this
implies ${\tilde A} = 0$. In dimensions
$n = 2 \pmod 4$ these conditions imply the
${\tilde A}$ is a multiple of the volume form
of the Clifford algebra. Therefore, if we
assume, as in Section~\ref{sec:eom} that
this possibility is removed, then we again
find that we require ${\tilde A} = 0$.

Generally, therefore, the constraints of higher
order in the Clifford algebra,
$({\tilde \Phi}, {\tilde \Psi})$,
tell us that the higher order variables,
$({\tilde A}, {\tilde \Pi})$, are redundant as far as the
dynamics of the theory are concerned.
Both the higher order constraints and
the higher order variables may therefore be
removed from the theory, and will not
appear again in this paper.
All of the spinorial variables and constraints left in the
theory are purely first and second order in the Clifford
algebra (from the
$\mathbb{C}\mathrm{l}_{n-1}$
point of view). We thus are
left with a theory with variables $(A_i, e_I^i)$,
conjugate momenta $(\Pi^i, \pi^I_i)$, and constraints
$(\Psi, \chi_{2i}, \chi_3, \Phi^i, \phi^I_i)$.

We should perhaps note that in dimension
$2 \pmod 4$, if we did not remove by
hand the unwanted terms from the theory,
then the division into first and second
class constraints given breaks down
in the case of the highest
order constraints. This leaves exactly
the extra freedom in the definition of the
connection given in Section~\ref{sec:eom}
and the corresponding extra freedom in the
conjugate momentum. What this extra freedom
corresponds to in gravitational terms is
not apparent, since the effect is spinorial
in nature and would not occur in a purely
tensorial approach.

\subsection{Minimal Theory}
\label{sec:min}
Assuming linear independence of the
vector fields $\mathbf{e}_I$ on $\Sigma$,
we decompose the remaining variables
$(A_i, \Pi^i)$ and constraints
$( \Psi , \Phi^i)$ using the densitised
$\gamma$-matrices ${\tilde\sigma}^i$ by defining
\begin{align*}
A_i{}^{j} &:=
\Tr \left[
A_i {\tilde\sigma}^j \right], & A_i{}^{jk} &:=
\Tr \left[
A_i {\tilde\sigma}^{[j}{\tilde\sigma}^{k]} \right],
\\
\pi^{ij} &:=
\Tr \left[
\pi^i {\tilde\sigma}^j \right],
&
\pi^{ijk} &:=
\Tr \left[
\pi^i {\tilde\sigma}^{[j}
{\tilde\sigma}^{k]} \right],
\end{align*}
and
\begin{align*}
\phi^{ij} &:=
\Tr \left[
\Phi^i {\tilde\sigma}^j \right],
&
\phi^{ijk} &:=
\Tr \left[
\Phi^i {\tilde\sigma}^{[j}{\tilde\sigma}^{k]} \right],
\\
\psi^i &:=
\Tr \left[
\Psi {\tilde\sigma}^i \right],
&
\psi^{ij} &:=
\Tr \left[
\Psi {\tilde\sigma}^{[i}
{\tilde\sigma}^{j]} \right].
\end{align*}
The arguments of the previous section
imply that, when we have eliminated
the higher order quantities from the
theory, we arrive at a theory with
variables $(A_i{}^j, A_i{}^{jk};
\pi^{ij}, \pi^{ijk})$, constraints
$(\phi^{ij}, \phi^{ijk}, \psi^i,
\psi^{ij}, \chi_{2i}, \chi_3)$
and Hamiltonian
\begin{equation}
H_T = - \int_{\Sigma}
A_{ti} \psi^i + A_{tij} \psi^{ij}
+ N^i \chi_{2i}
+ {\underline N} \chi_3
+ \int_{\Sigma} {\lambda}_{ijk} {\phi}^{ijk} +
{\lambda}_{ij} {\phi}^{ij}
+ \lambda_I^i \phi^I_i.
\label{ham}
\end{equation}

We now wish to consider the constraints
$( {\phi}^{ij}, \phi^I_i )$. It is
convenient to define linear
combinations of the first set of
constraints:
\[
{\phi}^i_I = q^{-1/2} e_{Ij} {\phi}^{ij}.
\]
It follows from the Poisson Bracket
relations that the families of constraints
$\{ {\phi}^i_I \}$ and $\{ {\phi}^I_i \}$
Poisson commute amongst themselves:
\begin{equation}
\left[ \phi_I^i (x), \phi_J^j (y) \right] = 0,
\qquad
\left[ \phi_i^I (x), \phi_j^J (y) \right] = 0,
\label{pb1}
\end{equation}
but that there are non-trivial Poisson Brackets
between the two families
\[
\left[ \phi_I^i (x), \phi^J_j (y) \right]
= {\dv} \, q^{1/2}
\left( \delta_I^J \delta_j^i -
e^J_j e^i_I \right) \delta (x, y).
\]
If $n \neq 2$, we can
invert the operator on the right hand
side of this equation to deduce that
\begin{equation}
\left( \delta^J_K \delta^k_j -
(n-2)^{-1} e_K^k e_j^J \right)
\left[ \phi_I^i (x),
\phi^K_k (y) \right]
= {\dv} \, q^{1/2} \,
\delta_I^J \, \delta_j^i \, \delta (x, y).
\label{pb2}
\end{equation}
Therefore we deduce
that, if $n \neq 2$, the matrix of Poisson Brackets of these
two families of constraints is of maximal
rank. Assuming linear independence of the
vector fields $\mathbf{e}_I$, we may therefore consider the
constraints $\phi^{ij}$ and $\phi^I_i$ as a conjugate
set of second-class constraints to be
removed from the theory. Constructing the
inverse of the matrix of Poisson Brackets,
and following the prescription of
equation~\eqref{dirbrac}, we define the Dirac Bracket
\begin{multline*}
\left[ f(y), g(z) \right]^* =
\left[ f(y), g(z) \right] -
\frac{1}{\dv} \int_{\Sigma}
d^{n-1}x \, q^{-1/2}
\left( \delta^J_I \delta^i_j
- (n-2)^{-1} e_I^i e_j^J \right)
\\
\times \left\{ \left[ f(y), \phi^I_i(x) \right]
\left[ \phi^j_J (x), g(z) \right]
-
\left[ f(y), \phi^j_J (x) \right]
\left[ \phi^I_i(x) , g(z) \right]
\right\}.
\end{multline*}
for any functions $f$ and $g$ on the phase space.
With the Poisson Brackets
redefined thus, the constraints $( \phi^I_i, \phi^{ij})$ have
vanishing Poisson Brackets with all functions on the phase
space. We are therefore free to impose the
constraints $\phi^I_i, \phi^{ij}$ as identities
on the theory:
\[
\pi^I_i = 0,
\qquad
\pi^{ij} = - {\dv} \, q \, q^{ij}.
\]
These identities may be looked on as completely removing
the variables $\pi^I_i$ and $\pi^{ij}$ from the theory.
When we define the Poisson Brackets of the reduced theory
with these variables removed as the Dirac Bracket above,
we find that the only non-trivial Poisson Brackets of the
remaining variables are
\begin{subequations}
\label{cc}
\begin{align}
\left[ A_i{}^j (x), A_k{}^l (y) \right] &=
\left( \delta_i^l A_k{}^j - A_i{}^l \delta_k^j \right) \delta (x, y),
\label{cc1}
\\
\left[ A_i{}^j (x), {\tilde\sigma}^k (y) \right] &=
\delta_i^k {\tilde\sigma}^j \delta (x, y),
\label{cc2}
\\
\left[ A_i{}^j (x), \pi^{klm} (y) \right] &=
\left( \delta_i^l \pi^{kjm} - \delta_i^m \pi^{kjl} \right) \delta (x, y),
\label{cc3}
\\
\left[ A_i{}^j (x), A_k{}^{lm} (y) \right] &=
\left( \delta_i^l A_k{}^{jm} - \delta_i^m A_k{}^{jl} \right) \delta (x, y),
\label{cc4}
\\
\left[ A_i{}^{jk} (x), {\pi}^{lmn} (y) \right] &=
- {\dv} \, q^2 \, \delta_i^l \,
\left( q^{jm} q^{kn} - q^{jn} q^{km} \right)
\delta (x, y),
\label{cc5}
\end{align}
\end{subequations}
where we have dropped the asterisks from the Dirac Bracket.
These Poisson Brackets are
the net effect of removing the
second-class constraints
$(\phi^I_i, \phi^{ij})$.
Equations~\eqref{cc1} and \eqref{cc2} imply that
$( A_i{}^j, {\tilde\sigma}^i)$
are not quite canonically conjugate variables,
but obey relations analogous to
variables $(q_i p^j, p^i)$ in ordinary mechanics,
as is to expected from the definition
of $A_i{}^j$.

Therefore, we have arrived at a
\lq\lq minimal\rq\rq\ version of
the general Hamiltonian theory with
canonical variables
\begin{equation}
( A_i{}^j, {\tilde\sigma}^i, A_i{}^{jk}, \pi^{ijk} ),
\label{minvars}
\end{equation}
which obey the Poisson Bracket relations above.
We have constraints
\begin{equation}
( \phi^{ijk}, \psi^i, \psi^{ij}, \chi_{2i}, \chi_3)
\label{mincons}
\end{equation}
and time evolution is generated by the Hamiltonian
\begin{equation}
H_T = - \int_{\Sigma}
A_{ti} \psi^i + A_{tij} \psi^{ij}
+ N^i \chi_{2i}
+ {\underline N} \chi_3
+ \int_{\Sigma} {\lambda}_{ijk} {\phi}^{ijk}.
\label{minham}
\end{equation}

The behaviour of this theory in
dimension $3$, or in dimension $4$
where we restrict to the
positive chirality spin-bundle, is quite
different from that in
higher dimensions, or in
dimension $4$ where we
work on the full spin-bundle.
In the next sections, we will
therefore first treat the
two special cases, and then in
Section~\ref{sec:general}
we consider the most general scenario.

\section{Ashtekar Variables in Dimensions $3$ and $4$}
\label{sec:3and4}

Here we summarise the main
simplifications that occur in
dimensions $3$ and $4$.
Since this material is, by now, standard
\cite{JS,S1,A2} we will be brief.

\subsection{Dimension $3$}

If the manifold $X$ is of dimension $3$, we may take the
$\gamma$-matrices to be proportional to the Pauli matrices
\[
\gamma^0 = - i \sigma^3, \qquad
\gamma^1 = - i \sigma^1, \qquad
\gamma^2 = - i \sigma^2.
\]
This means that
\[
\gamma^{AB} = \epsilon^{ABC} \gamma_C,
\]
where $\epsilon^{ABC} = \epsilon^{[ABC]}$
with $\epsilon^{012} = 1$ and internal
indices are raised and lowered using
the internal metric
$(\eta_{AB}) = (\eta^{AB}) = \delta_{AB}$.
Defining the two dimensional epsilon tensor
$\epsilon^{IJ} = \epsilon^{0IJ}$ then
\[
\Gamma^I = \gamma^{0I} = \epsilon^{IJ} \gamma_J.
\]
In this case there is no distinction
between pure elements of order
$1$ and $2$ in the Clifford algebra.
Therefore constraints of order $2$
drop out of the theory, with the only
constraints being the second-class constraints
$(\phi^{ij}, \phi^I_i)$ along with the
constraints $( \chi_1, \chi_{2i}, \chi_3 )$
which are first class.
Removing the second-class constraints as above,
we are left with a theory with
canonically conjugate variables
$( A_i, {\tilde\sigma}^j )$ and Hamiltonian
\begin{equation}
H_T = - \int_{\Sigma}
\Tr (A_{t} \chi_1)
+ N^i \chi_{2i}
+ {\underline N} \chi_3,
\label{3Dham}
\end{equation}
The constraint $\chi_1$ along with the equation of motion for the field
${\tilde\sigma}^i$ define the connection $\mathbf{A}$ to be the
spinorial image of the spin-connection. Constraints $\chi_{2i}$ and
$\chi_3$ along with the equation of motion for the connection then tell
us that this connection is flat. We therefore recover the usual
description of Ricci-flat geometry in $3$-dimensions.
The quantisation of this theory has been discussed
in some detail \cite{A2,W2}.

\subsection{Dimension $4$ restricted to $\mathbb{V}^+$}

We define a set of $4$-dimensional $\gamma$-matrices
\[
\gamma^0 = \epsilon \otimes 1, \qquad
\gamma^I = - i \sigma^1 \otimes \sigma^I,
\]
where $\epsilon = i \sigma^2 = \bigl( \begin{smallmatrix}
0&1\\-1&0 \end{smallmatrix} \bigr)$.
We define the volume element
\[
\omega = - \gamma^0 \gamma^1 \gamma^2 \gamma^3 = \sigma^3 \otimes
\mathrm{Id}.
\]
Restricting to $\mathbb{V}^+ = \{ \psi \in \mathbb{V}
| \omega \psi = \psi \}$, then the restricted generators are
\[
\Gamma^I = -i \sigma^I, \qquad
\Gamma^{IJ} = \epsilon^{IJK} \Gamma_K,
\]
where $\epsilon^{IJK} = \epsilon^{[IJK]}$ with $\epsilon^{123} = 1$.
In this case, only the self-dual part of the spin-connection
\[
A^+ = \frac{1}{2}
\left( A^{0I} + \frac{1}{2} \epsilon^{IJK} A_{JK} \right)
\Gamma_I
\]
appears, and the constraints $\phi^{ijk}$ are again
dropped from the theory. Removing the second-class constraints
$(\phi^{ij}, \phi^I_i)$ again leads to canonically
conjugate variables $(A^+_i, {\tilde\sigma}^i)$, and the
constraints of the theory are
\begin{align*}
\chi_1 &= D^{A^+} \cdot {\tilde\sigma},
\\
\chi_{2i} &= - \Tr \left( F^{A^+}_{ij} {\tilde\sigma}^j \right),
\\
\chi_3 &= \frac{1}{2} \Tr
\left( F^{A^+}_{ij} {\tilde\sigma}^i {\tilde\sigma}^j \right).
\end{align*}

This is the standard (Riemannian)
version of the Ashtekar theory
\cite{A1,JS,S1}.

\section{General dimension}
\label{sec:general}

The theory we derived in Section~\ref{sec:min} defined by
variables \eqref{minvars}, constraints \eqref{mincons} and
Hamiltonian~\eqref{minham} has extra
second class constraints in dimensions greater than $4$, or in
dimension $4$ without restricting to the positive spin-bundle.
We define the set of constraints
\[
\rho^i := |q|^{-1} q_{jk} \phi^{jki}.
\]
(In the notation of Section~\ref{sec:higher} this would
correspond to $\Lambda^1 [ \Tr \phi ]$ multiplied by $|q|^{-1}$).
It follows from the Poisson Bracket
relations \eqref{cc} that constraints
$\{ \psi^i \}$ and $\{ \rho^i \}$
Poisson commute amongst themselves
\[
\left[ \psi^i (x), \psi^j (y) \right] = 0,
\qquad
\left[ \rho^i (x), \rho^j (y) \right] = 0,
\]
but that there are non-trivial Poisson Brackets
\[
\left[ \psi^i (x), \rho^j (y) \right]
= - 2 {\dv} \, (n-2) \, q \, q^{ij}
\, \delta (x, y).
\]
It therefore follows that, if $n \neq 2$, the two families
of constraints $\{ \psi^i \}$, $\{ \rho^i \}$
are second-class, and their matrix of Poisson
Brackets is of maximal rank. Following the
Dirac procedure again, we deduce the new Dirac Bracket
\begin{multline*}
\left[ f(y), g(z) \right]^* =
\left[ f(y), g(z) \right] -
\frac{1}{2 (n-2) \dv}
\int_{\Sigma} d^{n-1}x \, q^{-1} \, q_{ij}
\\
\times \left\{ \left[ f(y), \psi^i(x) \right]
\left[ \rho^j (x), g(z) \right]
-
\left[ f(y), \rho^j (x) \right]
\left[ \psi^i (x) , g(z) \right]
\right\}.
\end{multline*}
for any functions $f$ and $g$ on the phase space.

In the notation of Section~\ref{sec:higher}, the constraints
$(\psi^i, \rho^i)$ impose that the $\Lambda^1 [\mathrm{Tr} \pi]$
momenta vanish, and that the variables
$\Lambda^1 [\mathrm{Tr} A]$ are defined by the equation
\begin{equation}
A_i{}^{ij} = - \frac{1}{2}
\Tr ( {\tilde\sigma}^j \partial_i {\tilde\sigma}^i ).
\label{aiijdef}
\end{equation}

Following Dirac's procedure, we therefore arrive at a
theory with dynamical variables
\[
( A_i{}^j, {\tilde\sigma}^i,
\Lambda^3 [\mathbf{A}], S_0^3 [\mathbf{A}],
\Lambda^3 [\pi], S_0^3 [\pi]),
\]
with the variables $A_i{}^{ij}$ of the old theory now being
derived quantities defined by equation~\eqref{aiijdef}.
The Poisson Bracket relations
\eqref{cc} imply that the variables
$( {\tilde\sigma}^i,
\Lambda^3 [\mathbf{A}], S_0^3 [\mathbf{A}],
\Lambda^3 [\pi], S_0^3 [\pi])$ Poisson commute with
the constraints $(\psi^i, \rho^i)$. Therefore, the Dirac
Brackets involving these quantities are identically equal
to the Poisson Brackets derived from the relations
\eqref{cc2}--\eqref{cc5}. The only remaining Dirac
Bracket we need consider is therefore
$\left[ A_i{}^j (x), A_k{}^l (y) \right]^*$.
Although the Poisson Brackets
of the constraints with $A_i{}^j$ do not vanish, the
Poisson Bracket of $A_i{}^j$ with the constraints
$\{ \rho^i \}$ vanishes weakly. Therefore, modulo constraints,
this Dirac Bracket is equal to the Poisson Bracket
\eqref{cc1}.
It is easily checked that the remaining constraints of the
theory, $( \psi^{ij}, \chi_{2i}, \chi_3,
\Lambda^3 [\phi], S_0^3 [\phi])$, are all first class.

This is often the most useful form of the theory to work with if
there are geometrical conditions one wishes to impose directly upon
the curvature $\mathbf{F}$ or the connection $\mathbf{A}$. This is
especially true if these geometrical conditions lead to the
constraints $\chi_{2i}$ and $\chi_3$ being satisfied automatically.
A particular example of such a simplification is in the study of
Ricci-flat Riemannian manifolds with reduced holonomy group. The
reduction in holonomy group may be attributed to the existence of
covariantly constant spinor fields \cite{LM}, which in turn leads
to restrictions on the curvature of the spin-connection. These
conditions may be integrated up, on a simply connected region, to
restrictions on the spin-connection and, in this case, it is
simplest to work with the full Hamiltonian theory~\eqref{minham}
\cite{G}.

Alternatively, we may proceed to eliminate the remaining parts of the
$\phi$ constraints from the theory. The constraints
$( \Lambda^3 [\phi], S_0^3 [\phi])$ impose that the remaining momenta,
$( \Lambda^3 [\pi], S_0^3 [\pi])$, vanish. Preservation of these
conditions serves to define the remaining parts of the spatial
connection $A_i{}^{jk}$ in terms of the vector fields $\mathbf{e}_I$,
in a way which is consistent with this connection being
identified with the (twice densitised)
$(n-1)$-dimensional spin-connection.

Given that the preservation of the remaining constraints
simply gives the definition of the remaining parts of the
connection, it is possible to look on these definitions
in themselves as extra constraints of the theory. These
constraints would then be canonically conjugate to the
remaining parts of the constraints $\phi^i$, and therefore
second-class. We can therefore remove all of these constraints
from the theory, imposing them as identities.
We are left with a theory with canonically conjugate variables
$(A_i{}^j, {\tilde\sigma}^k)$ along with constraints $(\psi^{ij},
\chi_{2i}, \chi_3)$. We must, however, rewrite these constraints in
terms of the canonical variables, replacing each part of the spatial
connection $A_i{}^{jk}$ by its expression in terms of
${\tilde\sigma}^i$. This leaves us with a theory with Hamiltonian
\[
H_T = - \int_{\Sigma}
A_{tij} \psi^{ij}
+ N^i \chi_{2i}
+ {\underline N} \chi_3,
\]
where we find that the constraints take the form
\begin{align*}
\psi^{ij} &= \frac{1}{2} \left( A^{ij} - A^{ji} \right),
\\
\chi_{2i} &= D_j A_i{}^j - D_i A_j{}^j,
\\
\chi_3 &= \frac{\dv}{4} \left[
\Tr \left( A^2 \right) - \Tr \left( A \right)^2 + s(\mathbf{q}) \right].
\end{align*}
In this equations, we have used the soldering form to construct
the twice densitised inverse metric with components
\[
| \mathrm{det} q| q^{ij} =
- \frac{\Tr ( {\tilde\sigma}^i {\tilde\sigma}^j )}{\dv},
\]
and $A^{ij} = q^{ik} A_k{}^j$. With this metric we then construct
the Levi-Civita connection $D$ and scalar curvature
$s(\mathbf{q})$. The formalism has therefore reduced to a spinorial
version of the standard ADM Hamiltonian theory. The field $A_i{}^j$
corresponds to a densitised version of the extrinsic curvature
$k_{ij}$, and the soldering forms ${\tilde\sigma}^i$ are a
densitised spinorial version of the vector fields $\mathbf{e}_I$.

The Poisson Bracket relations~\eqref{cc2} imply that all of the
constraints are first-class. Counting degrees of freedom, we have
$2(n-1)^2$ variables, $(A_i{}^j, {\tilde\sigma}^k)$, and
$\frac{1}{2}(n-1)(n-2)+(n-1)+1$ first-class constraints. Therefore,
we have $n(n-3)$ Hamiltonian degrees of freedom, as expected.

\section{Conclusion}

We have given a spinorial set of
Hamiltonian variables for General
Relativity which work in any
dimension greater than $2$.
Although, for simplicity, we have
restricted ourselves to Riemannian
signature, a similar analysis carries
through in any signature,
with appropriate minus signs.

In dimensions $0 \pmod 4$, the
theory can be reduced to the positive
chirality spin-bundle, but this
is not possible in dimensions $2 \pmod
4$ for algebraic reasons. It is
noticeable that the theory is very
different in the special cases
of $3$ dimensions and in $4$ dimensions
when restricted to the positive
spin-bundle. In these cases, the
constraint $\chi_1$ is first-class,
and the $\phi^i$ constraints
have no effect other than
setting $\pi^i = {\tilde\sigma}^i$.
In the more general case, the
constraints $\phi^i$ are conjugate
to parts of the $\chi_1$ constraints,
and we are forced to remove
part of the spatial part of the connection
from the problem. As in the
Palatini formalism, it is this
removal of the extra constraints
which leads to the apparent
non-polynomial nature of
remaining constraints \cite{A2}.
Whether it is useful to
reduce the theory to ADM
form seems to depend on the type
of problems we wish to tackle. If we
wish to impose geometrical conditions
directly on the connection or its
curvature, it seems more useful
to proceed without removing the
extra constraints first. This
may also be the more useful course
if one wishes to quantise the theory.

One obvious problem would be to
consider the extension of this approach
to canonical supergravity theories.
Certainly four-dimensional
minimal supergravity can be written
in Ashtekar-type form \cite{J}, and
recent attempts to find a unified
approach to the gravitational
field and the $3$-form potential
of $11$-dimensional supergravity seem
to suggest links with Ashtekar
variables \cite{MN}. It also seems
possible that our approach may be
useful in dimensions $0 \pmod 4$ when
one considers theories with chiral fermions.

A more geometrical problem mentioned earlier concerns the
description of Ricci-flat Riemannian metrics with reduced holonomy
group. The reductions of the holonomy group to Ricci-flat
K{\"a}hler (dimension $n=2k$), hyper-K{\"a}hler ($n=4k$), and
$\mathrm{G}_2$ ($n=7$) or $\mathrm{Spin}_7$ ($n=8$) may be
described in terms of the existence of covariantly constant
spinors of various types on a manifold \cite{LM}. An analysis of
this problem in light of the current formalism will be given
elsewhere \cite{G}.

\begin{ack}
This work was funded by the EPSRC.
\end{ack}

\end{document}